\documentstyle[12pt]{article}

\newcommand{\nc}{\newcommand}
\nc{\rnc}{\renewcommand}
\nc{\acs}{\arraycolsep}
\nc{\mc}{\multicolumn}
\nc{\bsk}{\baselineskip}
\nc{\vsp}{\vspace}
\nc{\hsp}{\hspace}
\nc{\stl}{\setlength}
\nc{\stc}{\setcounter}
\nc{\addl}{\addtolength}
\nc{\beq}{\begin{equation}}
\nc{\eeq}{\end{equation}}
\nc{\beqa}{\begin{eqnarray}}
\nc{\eeqa}{\end{eqnarray}}
\nc{\tfrac}[2]{\raisebox{.4ex}{\tiny $\frac{#1}{#2}$}}
\nc{\romlist}{ \setcounter{num1}{0}%
  \begin{list}{(\roman{num1})}{\usecounter{num1}} }
\nc{\arblist}{ \setcounter{num1}{0}%
  \begin{list}{(\arabic{num1})}{\usecounter{num1}} }
\nc{\alphlist}{ \setcounter{num2}{0}%
  \begin{list}{(\alph{num2})}{\usecounter{num2}} }
\nc{\bullist}{\begin{list}{$\bullet$}{ }}
\nc{\nr}{\\ \hline}
\nc{\hrl}{{\center \stl{\unitlength}{\textwidth}
 \begin{picture}(1,0)  \put(0,0){\line(1,0){1}}
 \end{picture} \vsp{.001\bsk} }}
\nc{\cents}{{\scriptsize$\mbox{\rm C}\!\!\!\mbox{\raisebox{.2ex}%
{$|$}}\,\,\,\,$}}
\nc{\figsp}[5]{\begin{figure}[#1] \vsp{#2} \caption[#4]{#3}
\label{#5} \vsp{2\bsk} \end{figure}}
\nc{\fig}{\figsp{tbp}}
\nc{\figb}{\figsp{b}}
\nc{\figh}{\figsp{h}}
\nc{\llist}{\begin{list}{}{} \stl{\labelsep}{.4in}}
\nc{\lit}[2]{
 \item[\raggedright #1]{#2}}
\nc{\lbit}[2]{
 \item[\raggedright\bf #1]{#2}}
\nc{\lemit}[2]{
 \item[\raggedright\em #1]{#2}}
\nc{\lbemit}[2]{
 \item[\raggedright\bf\em #1]{#2}}
\nc{\clst}[1]{\stl{\coltwo}{\textwidth}
\addl{\coltwo}{-#1} \addl{\coltwo}{-5.56ex} \newline
\begin{tabular}{p{#1}p{\coltwo}} \citem{}{}}

\nc{\citem}[2]{{\raggedright \bf #1} & #2 \\ }
\nc{\cemitem}[2]{{\raggedright \em #1} & #2 \\ }
\nc{\cbemitem}[2]{{\raggedright \bf \em #1} & #2 \\ }
\nc{\cend}{\citem{}{} \end{tabular}
\mbox{}
}
\nc{\SSP}{{\rm \hsp{.4in}}}
\nc{\SSPP}{{\rm \hsp{.2in}}}
\nc{\ds}{\displaystyle}
\nc{\tx}{\textstyle}
\nc{\scst}{\scriptstyle}
\nc{\sscst}{\scriptscriptstyle}
\nc{\prt}{\partial}
\nc{\fr}{\frac}
\nc{\lf}{\left}
\nc{\rt}{\right}
\nc{\la}{\langle}
\nc{\ra}{\rangle}
\nc{\V}{\vec}
\nc{\str}{\stackrel}
\nc{\ovl}{\overline}
\nc{\ul}{\underline}
\nc{\ovb}{\overbrace}
\nc{\ub}{\underbrace}
\nc{\wh}{\widehat}
\nc{\B}{\bar}
\nc{\D}{\dot}
\nc{\C}{\cdot}
\nc{\dd}{\ddot}
\nc{\tl}{\tilde}
\nc{\ha}{\hat}
\nc{\nn}{\nonumber}
\nc{\app}{\approx}
\nc{\al}{\alpha}
\nc{\RA}{\rightarrow}
\nc{\LRA}{\leftrightarrow}
\nc{\SRA}{\SSP\rightarrow\SSP}
\nc{\SSRA}{\SSPP\rightarrow\SSPP}
\nc{\dg}{\dagger}
\nc{\vp}{\varphi}
\nc{\ve}{\varepsilon}
\nc{\Dl}{\Delta}
\nc{\dl}{\delta}
\nc{\gm}{\gamma}
\nc{\Gm}{\Gamma}
\nc{\ep}{\epsilon}
\nc{\sg}{\sigma}
\nc{\Sg}{\Sigma}
\nc{\ua}{\uparrow}
\nc{\da}{\downarrow}
\nc{\lam}{\lambda}
\nc{\Lam}{\Lambda}
\nc{\eql}[1]{\parbox{#1\textwidth}}
\nc{\eqm}[1]{\makebox[#1\textwidth][l]}
\nc{\enu}[1]{\mbox{\hspace{.4in}(\theequation.#1)}}
\nc{\son}{\\ \\ \ds}
\nc{\stw}{\\ & \\ \ds}		
\nc{\sth}{\\ & & \\ \ds}
\nc{\sfo}{\\ & & & \\ \ds}
\nc{\sfi}{\\ & & & & \\ \ds}
\nc{\A}{& \ds}
\nc{\bbr}{\lf\{\rule[-1.5ex]{0in}{0.01in}\rt.}
\nc{\hf}{\fr{1}{2}}
\nc{\mhf}{\mbox{\footnotesize$\hf$}}
\nc{\dv}{\/!}
\nc{\sint}{\int\!\!}
\nc{\dint}{\int\!\!\sint}
\nc{\tint}{\int\!\!\dint}               
\nc{\qint}{\int\!\!\tint}
\nc{\Pd}[2]{\fr{\prt #1}{\prt #2}}
\nc{\Pdt}[1]{\Pd{#1}{t}}
\nc{\Pdx}[1]{\Pd{#1}{x}}
\nc{\Pdy}[1]{\Pd{#1}{y}}
\nc{\Pdz}[1]{\Pd{#1}{z}}           	
\nc{\Pdr}[1]{\Pd{#1}{r}}
\nc{\Pds}[1]{\Pd{#1}{s}}
\nc{\Dv}[2]{\fr{d#1}{d#2}}
\nc{\Dvt}[1]{\Dv{#1}{t}}
\nc{\Dvx}[1]{\Dv{#1}{x}}
\nc{\Dvy}[1]{\Dv{#1}{y}}
\nc{\Dvz}[1]{\Dv{#1}{z}}
\nc{\Dvr}[1]{\Dv{#1}{r}}
\nc{\Drs}[1]{\Dv{#1}{s}}
\nc{\inpp}[3]{\la #1| #2| #3\ra}
\nc{\inp}[2]{\inpp{#1}{#2}{#1}}
\nc{\rb}[1]{| #1\ra}
\nc{\lb}[1]{\la#1|}
\nc{\dtpp}[2]{\lb{#1}\rb{#2}}		
\nc{\dtp}[1]{\dtpp{#1}{#1}}
\nc{\otpp}[2]{\rb{#1}\lb{#2}}
\nc{\otp}[1]{\otpp{#1}{#1}}
\rnc{\L}{{\cal L}}                      
\nc{\lapp}{\mbox{\raisebox{-.6ex}{$\,\stackrel{\textstyle <}{\sim}\,$}}}
\nc{\gapp}{\mbox{\raisebox{-.6ex}{$\,\stackrel{\textstyle >}{\sim}\,$}}}
\newcounter{num1} \newcounter{num2}  
\newlength{\coltwo}
\nc{\als}{\fr{\al_s(Q^2)}{2\pi}}
\nc{\gpx}{g_1^p(x,Q^2)}
\nc{\gpz}{g_1^p(z,Q^2)}
\nc{\muq}{\lf(\fr{\mu^2}{Q^2}\rt)}
\nc{\xy}{(\fr{x}{y})}
\nc{\ASQ}{\al_s(Q^2)}
\nc{\Li}{{\rm Li}_2}
\nc{\dqx}{\Dl q_i(x,Q^2)} \nc{\dqy}{\Dl q_i(y,Q^2)}
\nc{\dQ}{\Dl q_i(Q^2)}
\nc{\dgx}{\Dl g(x,Q^2)} \nc{\dgy}{\Dl g(y,Q^2)}
\nc{\dG}{\Dl g(Q^2)}
\nc{\xq}{(x,Q^2)} \nc{\yq}{(y,Q^2)}
\nc{\Tt}{\tl{t}} \nc{\Ts}{\tl{s}} \nc{\Tu}{\tl{u}}
\nc{\Hs}{\ha{s}} \nc{\Ht}{\ha{t}} \nc{\Hu}{\ha{u}}
\nc{\Hsg}{\hat{\sg}}
\nc{\GeV}{\mbox{\rm GeV}}
\nc{\sS}{\!\not{\!s}}
\nc{\pS}{\!\not{\!p}}  \nc{\kS}{\!\not{\!k}}
\nc{\poS}{\!\not{\!p}_1}  \nc{\pwS}{\!\not{\!p}_2}
\nc{\ptS}{\!\not{\!p}_3}  \nc{\pfS}{\!\not{\!p}_4}
\nc{\AS}{\!\not{\!\!A}}  \nc{\ASS}{\!\not{\!\!A}^*}
\nc{\BS}{\!\not{\!\!B}}  \nc{\BSS}{\!\not{\!\!B}^*}
\nc{\Tr}{\mbox{\rm Tr}}
\nc{\pT}{p_T}
\nc{\xT}{x_T}
\nc{\AoS}{\!\not{\!\!A}_1}  \nc{\AoSS}{\!\not{\!\!A}_1^*}
\nc{\AwS}{\!\not{\!\!A}_2}  \nc{\AwSS}{\!\not{\!\!A}_2^*}
\nc{\BoS}{\!\not{\!\!B}_1}  \nc{\BoSS}{\!\not{\!\!B}_1^*}
\nc{\BwS}{\!\not{\!\!B}_2}  \nc{\BwSS}{\!\not{\!\!B}_2^*}
\nc{\aS}{\!\not{\!a}}  \nc{\bS}{\!\not{\!b}}
\nc{\aoS}{\!\not{\!a}_1}  \nc{\boS}{\!\not{\!b}_1}
\nc{\awS}{\!\not{\!a}_2}  \nc{\bwS}{\!\not{\!b}_2}
\nc{\anS}{\!\not{\!a}_n}  \nc{\bnS}{\!\not{\!b}_n}
\nc{\anpoS}{\!\not{\!a}_{n+1}}  \nc{\bnpoS}{\!\not{\!b}_{n+1}}
\nc{\anmoS}{\!\not{\!a}_{n-1}}  \nc{\bnmoS}{\!\not{\!b}_{n-1}}
\nc{\refi}[1]{$^{\,\mbox{\scriptsize \ref{#1}}}$}
\nc{\refii}[2]{$^{\,\mbox{\scriptsize \ref{#1},\ref{#2}}}$}
\nc{\refiii}[3]{$^{\,\mbox{\scriptsize \ref{#1},\ref{#2},\ref{#3}}}$}
\nc{\refr}[2]{$^{\,\mbox{\scriptsize \ref{#1}--\ref{#2}}}$}

\renewenvironment{thebibliography}[1]
  { \begin{list}{[\arabic{enumi}]}
    {\usecounter{enumi} \setlength{\parsep}{0pt}
     \setlength{\itemsep}{3pt} \settowidth{\labelwidth}{[#1]}
     \sloppy
    }}{\end{list}}

\newenvironment{thecaptions}[1]
  { \begin{list}{\arabic{enumi}.}
    {\usecounter{enumi} \setlength{\parsep}{0pt}
     \setlength{\itemsep}{3pt} \settowidth{\labelwidth}{#1.}
     \sloppy
    }}{\end{list}}

\begin{document}

\pagestyle{empty}

\hfill {\bf McGill-95/48}

\hfill {\bf hep-ph/9511217}

\hfill August 1995

\vsp{1cm}

\begin{center} \begin{Large} \begin{bf}
QCD Corrections to Spin Dependent Drell-Yan and a Global
Subtraction Scheme
\end{bf} \end{Large} \end{center}
\vglue 0.35cm
{\begin{center}
 B.\ Kamal
\end{center}}
\parbox{6.4in}{
{\it Department of Physics, McGill University, Montreal,
Qu\'{e}bec, Canada, H3A 2T8}
}
\begin{center}
\vglue 1.0cm
\begin{bf} ABSTRACT \end{bf}
\end{center}
{
 \noindent
We present QCD corrections to the Drell-Yan process in the
transversely polarized, longitudinally polarized and unpolarized
cases. The analytical results are presented in a form valid
for all $n$-dimensional regularization schemes. A universal
mass factorization scheme is presented in which the results
reduce to those of dimensional reduction. The connection between
the parton distributions and fragmentation functions of
dimensional reduction and those of
dimensional regularization is
elucidated in a simple manner.
Numerical results are presented for proton-proton
collisions at energies relevant to RHIC (Relativistic Heavy
Ion Collider). The perturbative stability of the transverse and
longitudinal asymmetries is investigated.
}

\vsp{.15in}

\vsp{1in}
\begin{center}
(To appear in Phys.\ Rev.\ {\bf D})
\end{center}

\newpage

\pagestyle{plain}
\setcounter{page}{1}

\begin{center}\begin{large}\begin{bf}
I. INTRODUCTION
\end{bf}\end{large}\end{center}
\vglue .3cm

The unpolarized Drell-Yan process has been studied rather extensively
in the literature, including ${\cal O}(\al_s)$ \cite{AEM,Kub}
and ${\cal O}(\al_s^2)$ \cite{Hamb}
corrections. As well, the ${\cal O}(\al_s)$ corrections to the corresponding
longitudinally polarized \cite{Rat}
and transversely polarized processes
\cite{VW,PLB} have been studied.
What was still lacking is a unified picture for
dealing with the polarized processes. The basic problem is the
ambiguity associated with defining the $\gm_5$ matrix,
or $\ve^{\mu\nu\lam\rho}$ tensor, in $n$ dimensions;
both of these objects arise in polarized processes.  For
unpolarized QCD processes, {\em dimensional regularization} (DREG) preserves
all the necessary invariances and symmetries to do calculations to any
order in $\al_s$. Hence DREG is the most commonly used regularization
for QCD. The ambiguity associated with the continuation of the
$\gm_5$ matrix makes it impossible to uniquely define higher order
corrections (HOC) for polarized processes using DREG. Various prescriptions
are available, but problems with either mathematical or physical
consistency generally arise. As a result, another $n$-dimensional
scheme, {\em dimensional reduction} (DRED) may be used. This scheme
avoids the $\gm_5$ problem, although it requires certain
ultraviolet (UV) counterterms which are the same in both
unpolarized and polarized processes and may be unambiguously
determined.

In this paper, we present analytical results for unpolarized
and (both longitudinally and transversely) polarized
Drell-Yan in a form valid for all $n$-dimensional schemes.
For the polarized case, the ambiguity (or scheme dependence)
in the DREG results is parameterized by the ambiguity in the
polarized $n$-dimensional split functions. As well, we present
numerical results for $p$-$p$ collisions relevant to the
Relativistic Heavy Ion Collider (RHIC).

We go further to show that for a wide class of subprocesses,
including the one-loop corrections to Drell-Yan and deep-inelastic
scattering, DRED is simply equivalent to a particular mass factorization
scheme in DREG. We call this scheme $\ovl{\rm MS}_\ve$ (or
${\rm MS}_\ve$) since it involves subtracting the $\ve$-dimensional
part of the $n$-dimensional Altarelli-Parisi split functions,
where $n=4-2\ve$.
As a consequence, the final results in the $\ovl{\rm MS}_\ve$
scheme are regularization scheme independent within the
$n$-dimensional schemes. The final result is equivalent to that
obtained in DRED and all ambiguities associated with the continuation
of the $\gm_5$ matrix are subtracted via the $n$-dimensional
split functions.

We will also show the connection between the DRED  parton
distributions and fragmentation functions and those of DREG
in a simple manner.
More specifically, we show how to convert existing DREG
distributions into ones suitable for use with
cross sections determined using DRED. This is important since DRED
is equivalent to 4-dimensional helicity amplitude techniques which
considerably simplify perturbative calculations. We may thus
calculate new unpolarized cross sections using DRED or
helicity amplitudes and then simply convolute them with the
DRED distributions obtained from  well-known unpolarized DREG
parton distributions and fragmentataion functions.

Similar conclusions (for unpolarized processes) may be obtained
in the approach of \cite{Korner}, which converts DRED cross
sections into DREG ones by considering differences in the Lagrangians
and using fictitious $\ve$-scalars to calculate the differences
in the cross sections. Transition rules between the two schemes are
also given in \cite{Kun}.
Here, we take a simpler and more phenomenological
approach, investigating how the scheme dependences arise in
the Feynman graphs.
The connection between the two schemes is simply the relation
between the distributions of the respective schemes.
This allows for easy interpretation and
extension to a wide class of processes.
We also explicitly consider polarized observables, unlike
\cite{Korner,Kun}.

\vglue 1cm
\begin{center}\begin{large}\begin{bf}
II. $n$-DIMENSIONAL REGULARIZATION SCHEMES
\end{bf}\end{large}\end{center}
\vglue .3cm

There are two parts to the dimensional continuation: the
continuation of the {\em momenta} and the continuation of
all other {\em tensor structures} (i.e. gamma matrices). The
continuation of the momenta is unique, but there are various
methods for continuing the tensors. The choice of the latter
defines which dimensional method is being used.

\noindent {\em Continuation of the Momenta}

For the continuation of the momenta, all momenta and phase
spaces are continued to $n$ dimensions \cite{HV,Bol}. The phase space
integrals are continued by generalizing integer dimensional
integrals to non-integer dimensions. Consequently, all  loop
integrals can be reduced, using Feynman parameters, to the
fundamental integral
\beq
\label{e1}
\sint \fr{d^nq}{(2\pi)^n} \fr{(q^2)^r}{(q^2-C)^m}
= \fr{i(-1)^{r-m}}{(4\pi)^{n/2}\Gm(n/2)} C^{r-m+n/2}
B(r+n/2,m-r-n/2),
\eeq
(see, for example \cite{Marc}) with $m>0$, $r\geq 0$ and
$B$ the Euler beta function. As well, defining $n=4-2\ve$
($n'=4-2\ve'$) with $\ve<0$ ($\ve'>0$) we see that $\ve'$ is
required for UV divergent integrals and $\ve$ for infrared
(IR) divergent ones, initially. Then we must continue to
\beq
\ve'=\ve
\eeq
since we can only work in one dimension at one time. From
(\ref{e1}) it follows that massless self-energy insertions on
massless external lines vanish. This means that on-shell
wavefunction renormalization is trivial when all particles are
massless. Hence, in the absence of coupling renormalization
(i.e.\ gluon self-energies), effectively no UV renormalization
is required.

\noindent {\em Continuation of the Tensors (DREG)}

In DREG, one continues the metric tensor and the gamma matrices
to $n$ dimensions. Letting $g_n^{\mu\nu}$ denote the $n$-dimensional
metric tensor, we have the relations
\beq
g_n^{\mu\nu}g^n_{\mu\nu}=n, \SSP \gm^\mu\gm^\nu + \gm^\nu\gm^\mu
= 2 g_n^{\mu\nu}.
\eeq
As well, the usual convention is to take $2-2\ve=n-2$ helicity
states when averaging over initial gluons/photons.
This is related to the continuation of the helicity sum rule
\beq
\sum_{\lam} A^{\mu}(p,\lam) A^{*\nu}(p,\lam) \RA - g_n^{\mu\nu}.
\eeq
Here $A^{\mu}(p,\lam)$ is the gluon/photon polarization vector
for gluon/photon momentum $p$ and helicity $\lam$.
Different conventions simply amount to finite
renormalizations of the parton distributions (which we will see
arise from differences in the $n$-dimensional split functions).

There exist two popular methods for continuing the $\gm_5$ matrix
($\ve^{\mu\nu\lam\rho}$) tensor to $n$ dimensions: the anticommuting-%
$\gm_5$ scheme \cite{Chan} (see also \cite{APC}
concerning $\ve^{\mu\nu\lam\rho}$)
and the HVBM scheme \cite{HV,BM}.

In the anticommuting-$\gm_5$ scheme, we use the relations
\beq
\gm_5\gm_{\mu} = - \gm_\mu\gm_5, \SSP \gm_5^2=-1.
\eeq
If traces with only one $\gm_5$ occur though, there are known
mathematical inconsistencies\cite{BM}.

In the HVBM scheme, we formally take $n>4$ (with regards to the
tensor algebra) and keep the $\gm_5$ and $\ve^{\mu\nu\lam\rho}$
in 4 dimensions so that
\beq
\label{e5}
\mbox{$\{\gm_5,\gm_{\mu}\} = 0$: $\mu\leq 4$,
\SSP $[\gm_5,\gm_{\mu}]=0$: $\mu>4$},
\eeq
which follows from the definition
\beq
\gm_5 = \fr{i}{4!} \ve_{\mu_1\mu_2\mu_3\mu_4}
\gm^{\mu_1}\gm^{\mu_2}\gm^{\mu_3}\gm^{\mu_4}
\eeq
where
\beq
\ve_{\mu_1\mu_2\mu_3\mu_4}=0, \SSPP \mu_i > 4;
\eeq
otherwise, it is the usual Levi-Civita tensor.

This scheme is mathematically consistent,
but cumbersome.
Physically, it has the problem that the non-anticommuting $\gm_5$
leads to non-conservation of helicity of massless fermions in
a minimal subtraction scheme like $\ovl{\rm MS}$ \cite{Vog}.

\noindent{\em Dimensional Reduction}

Dimensional reduction \cite{Sie}
is perhaps the simplest of all the dimensional
methods. It was originally introduced because DREG violates
supersymmetry.
As will become obvious, it is also manifestly
mathematically consistent. The idea is simple; all $\gm$-matrices
and tensors are taken to be 4-dimensional, and formally $n<4$. This
implies that the components of all momenta between $n$ and 4 must
vanish. We have the following contraction identities
\beq
g^{\mu\nu}g_{\mu\nu} = 4, \SSPP g_n^{\mu\nu}g^n_{\mu\nu} =
g^{\mu\nu}g^n_{\mu\nu} = n
\eeq
and the usual 4-dimensional relations like
\beq
\gm^{\mu}\gm^{\nu} + \gm^{\nu}\gm^{\mu} = 2 g^{\mu\nu}.
\eeq
It is also useful to define
\beq
\label{gmep}
\gm_{\ve}^{\mu} \equiv \gm_{\nu}(g^{\mu\nu}-g_n^{\mu\nu}).
\eeq

This method is particularly simple for the calculation of tree
graphs (i.e.\ graphs not involving loops) since the traces are
equal to their 4-dimensional counterparts, implying gauge
invariance. One may thus use 4-dimensional helicity amplitude
methods, for instance. Then the phase space integrals are carried out in
$n$ dimensions, providing and IR regulator. As well, the anticommuting
$\gm_5$ implies helicity conservation of massless fermions.

The only subtlety comes from the fact that the virtual momentum
integrations generate the tensor $g_n^{\mu\nu}$, which is
generally contracted with 4-dimensional $\gm$-matrices.
This can lead to a term $\sim\gm_{\ve}^{\mu}$ which
gives the incorrect Lorentz structure and must be removed by a
counterterm. In \cite{Hvy}
 the counterterm for the quark-$\gm$($Z$)
vertex was presented. It is (working in the Feynman gauge)
\beq
\label{e11}
\gm^\mu \RA - C_F \fr{g^2}{(4\pi)^2} \fr{1}{\ve}\gm_\ve^\mu,
\eeq
with $C_F=4/3$
(i.e.\ the Feynman rule for the counterterm is obtained by
making the above substitution in the usual rule). For the
lepton-$\gm$($Z$) vertex, we use (\ref{e11}) with
\beq
C_F g^2 \RA e^2.
\eeq

Throughout, we consider (\ref{e11}) to be a Feynman rule for
DRED. For the type of processes considered here, (\ref{e11})
is the only counterterm required to make DRED physically consistent.

\vglue 1cm
\begin{center}\begin{large}\begin{bf}
III. ANALYTICAL RESULTS FOR  DRELL-YAN
\end{bf}\end{large}\end{center}
\vglue .3cm

We will first consider the unpolarized and longitudinally polarized
cases, then the transversely polarized case. We have the
general process
\beq
\label{e13}
A(P_1,\lam_A) + B(P_2,\lam_B) \RA l^-(p_3) + l^+(p_4) + X,
\eeq
where $\lam_A$, $\lam_B$ denote the helicities of hadrons $A$, $B$.
The unpolarized and longitudinally polarized cross sections are defined,
respectively, by
\beq \label{xsct}
\sg \equiv \fr{1}{2} [\sg(+,+)+\sg(+,-)],
\SSPP \Dl \sg \equiv \fr{1}{2} [\sg(+,+)-\sg(+,-)]
\eeq
in the notation $\sg(\lam_A,\lam_B)$.

The general $2\RA 2$ [$2\RA 3$] subprocess contributing to (\ref{e13})
has the form
\beq
\label{e15}
a(p_1,\lam_1) + b(p_2, \lam_2) \RA \gm^*(q) + [c(k)]
\RA l^-(p_3) + l^+(p_4) + [c(k)]
\eeq
for general partons $a$, $b$, $c$.

Firstly, we define the process-level invariants
\beq
S=(P_1+P_2)^2, \SSPP M^2=(p_3+p_4)^2, \SSPP \tau = M^2/S.
\eeq
In the parton model, we have
\beq
p_1 = x_a P_1,\SSP p_2=x_b P_2.
\eeq
Hence we may define the subprocess invariants
\beq
s=(p_1+p_2)^2 = x_ax_bS, \SSPP w=\fr{M^2}{s} =
\fr{M^2}{Sx_ax_b} = \fr{\tau}{x_ax_b}.
\eeq
The unpolarized [polarized] momentum distributions are given
by
\beq
[\Dl] F_{i/I} (x_i, M_f^2) = x_i [\Dl] f_{i/I} (x_i, M_f^2)
\eeq
where the $[\Dl]f_{i/I}$ are the unpolarized [polarized] parton densities
for parton $i$ in hadron $I$,
evaluated at factorization energy scale $M_f^2$.

The parton model expression for the Drell-Yan cross section
corresponding to (\ref{e13}) is then
\beq \label{e20}
[\Dl] \fr{d\sg_{AB}}{dM^2} = \sum_{ab}
\int_{\tau}^1\!\! \fr{dx_a}{x_a} \int_{w_1}^1 \fr{dw}{w}
[\Dl] F_{a/A}(x_a,M_f^2) [\Dl] F_{b/B}(x_b,M_f^2)
[\Dl] \fr{d\ha{\sg}_{ab}}{dM^2}
\eeq
where
\beq
w_1 = \tau/x_a, \SSP x_b = w_1/w
\eeq
and $[\Dl]\ha{\sg}_{ab}$ is the unpolarized [polarized] subprocess
cross section corresponding
to (\ref{e15}). We must consider the subprocesses $a=q$, $b=\B{q}$,
$c=g$ and $a=q$, $b=g$, $c=q$, which are symmetric under
$a\LRA b$ and $q\LRA \B{q}$ as far as $[\Dl]d\ha{\sg}_{ab}/dM^2$
is concerned.

Define the unintegrated leptonic tensor as
\beqa
\nn L_{\rm DRED}^{\al\beta} &=& \mu^{2\ve} e^2 [p_3^\al p_4^\beta
+ p_4^\al p_3^\beta - (M^2/2)g^{\al\beta}] \\
L_{\rm DREG}^{\al\beta} &=& \mu^{2\ve} e^2 [p_3^\al p_4^\beta
+ p_4^\al p_3^\beta - (M^2/2)g_n^{\al\beta}]
\eeqa
where the arbitrary mass scale $\mu^{2\ve}$ arises from the
$n$-dimensional coupling $e^2\RA e^2\mu^{2\ve}$. Furthermore,
define the integrated leptonic tensor as
\beq
{\cal L}^{\al\beta} = \sint \fr{d^{n-1}p_3}{(2\pi)^{n-1}2p_{3,0}}
\fr{\dl[(q-p_3)^2]}{M^2} L^{\al\beta}.
\eeq
One finds
\beq
{\cal L}_{\rm DRED}^{\al\beta} = \mu^{2\ve} e^2 \fr{\pi^\ve}{2^n}
\fr{(q^2)^{-\ve}}{\pi^2}
\fr{\Gm(1-\ve)}{\Gm(1-2\ve)} \fr{1}{(3-2\ve)(1-2\ve)}
\lf[ (1-\ve) \fr{q^\al q^\beta}{q^2} + \fr{g_n^{\al\beta}}{2}
- (3-2\ve) \fr{g^{\al\beta}}{2} \rt].
\eeq
The corresponding DREG tensor is obtained by replacing
$g^{\al\beta}\RA g_n^{\al\beta}$. This gives
\beq
{\cal L}_{\rm DREG}^{\al\beta} = \mu^{2\ve} e^2 \fr{\pi^\ve}{2^n}
\fr{(q^2)^{-\ve}}{\pi^2}
\fr{\Gm(1-\ve)}{\Gm(1-2\ve)} \fr{(1-\ve)}{(3-2\ve)(1-2\ve)}
\lf[  \fr{q^\al q^\beta}{q^2} - g_n^{\al\beta} \rt].
\eeq
The part $\sim q^\al q^\beta$ does not contribute to the cross
section, as can be seen from gauge invariance. Hence the
integrated leptonic tensor is effectively a constant. Nonetheless,
we keep all the terms for completeness.

We may then define the unpolarized [polarized] subprocess
hadronic tensor $[\Dl] W_{ab}^{\al\beta}$
through the unpolarized [polarized] subprocess squared Feynman amplitude
\beq
[\Dl] |M|_{ab}^2 \equiv \fr{1}{M^4} L_{\al\beta}
[\Dl] W_{ab}^{\al\beta},
\eeq
where $|M|^2_{ab}$, $\Dl |M|^2_{ab}$
are defined analogously to (\ref{xsct}).
Having done so, we may write the $2\RA 2$ phase space as
\beq
[\Dl] \fr{d\ha{\sg}_{ab,2\RA 2}}{dM^2} = \fr{1}{M^4}
\lf( 16 \pi \fr{\dl(1-w)}{M^2} [\Dl] W_{ab,2\RA 2}^{\al\beta} \rt)
{\cal L}_{\al\beta}.
\eeq
Similarly, for the $2\RA 3$ phase space,
\beq
[\Dl] \fr{d\ha{\sg}_{ab,2\RA 3}}{dM^2} = \fr{1}{M^4}
\lf[ \fr{2w}{\pi} \fr{\pi^{\ve}M^{-2\ve}}{2^{1-2\ve}}
\fr{w^\ve(1-w)^{1-2\ve}}{\Gm(1-\ve)} \int_0^1\!\! dy y^{-\ve}
(1-y)^{-\ve}
[\Dl] W_{ab,2\RA 3}^{\al\beta} {\cal L}_{\al\beta} \rt],
\eeq
where $y=(1+\cos\theta)/2$ and $\theta$ is the angle between
$p_1$ and $k$ in the $p_1$, $p_2$ c.m..
This is all we need to calculate $[\Dl] d\sg_{AB}/dM^2$.

In order to present $[\Dl] d\ha{\sg}_{ab}/dM^2$ in a form
valid for all $n$-dimensional schemes, we must first give the
general form of the $n$-dimensional split functions $P_{ij}^n(z)$,
related to the probability of parton $j$ splitting into a
collinear parton $i$ having momentum fraction $z$, plus an
arbitrary final state carrying the rest of the momentum.

We may write
\beq
P_{ij}^n(z,\ve) = P_{ij}^<(z,\ve) + \dl(1-z) P_{ij}^\dl(\ve),
\eeq
with
\beq
P_{ij}^<(z,\ve) = P_{ij}^{<,4}(z) + \ve P_{ij}^{<,\ve}(z)
\eeq
and
\beq
P_{ij}^\dl(\ve) = P_{ij}^{\dl,4} + \ve P_{ij}^{\dl,\ve}.
\eeq
In DRED, $P_{ij}^{<,\ve}(z)$ and $P_{ij}^{\dl,\ve}$ are zero.
In other words
\beq
P_{ij}^{\rm DRED}(z) = P_{ij}^4(z),
\eeq
where $P_{ij}^4(z)$ is the usual 4-dimensional  split function.

One might wonder how to determine the $P_{ij}^n(z,\ve)$. It is
done in the same way as for the 4-dimensional case, but keeping the
terms of ${\cal O}(\ve)$.

We can make this clearer by considering
the $2\RA m$ process (all particles massless)
\beq
a(p_a) + b(p_b) \RA c_1(k_1) + c_2(k_2) + \cdots + c_m(k_m).
\eeq
When $k_1$ is collinear with $p_a$, we have (at one-loop)
\beq
\label{e34}
[\Dl] |M|^2_{2\RA m}(ab\RA c_1\cdots c_m) \sim
\sum_d \fr{[\Dl] P_{da}^<(z,\ve)}{p_a\C k_1}
[\Dl] |M|^2_{2\RA m-1}(db\RA c_2 \cdots c_m),
\eeq
where $a\RA d + c_1$ and
\beq
k_1 \approx (1-z)p_a \RA p_d \approx z p_a, \SSP z<1,
\eeq
with  $\Dl P_{ij}$ being the corresponding polarized split function.
Then $P_{ij}^\dl$ is determined using probability and momentum
conservation, and it only appears when $i=j$.
Also, by definition $\Dl P_{ij}^\dl = P_{ij}^\dl$.

In this paper, we will need $[\Dl] P_{qq}(z)$ and $[\Dl] P_{qg}(z)$.
In 4 dimensions \cite{AP}
\beqa
\nn
\Dl P_{qq}^4(z) &=& P_{qq}^4(z) = C_F \lf[ \fr{1+z^2}{(1-z)_+}
+ \fr{3}{2} \dl(1-z) \rt]
= C_F \lf[ \fr{2}{(1-z)_+} -1 -z + \fr{3}{2} \dl(1-z) \rt]
\\
P^4_{qg}(z) &=& \fr{1}{2} (1-2z+2z^2),
\SSPP \Dl P^4_{qg}(z) = z-\fr{1}{2}.
\eeqa
For the unpolarized case in DREG, the $\ve$-dimensional parts are
unique \cite{ERT}
\beqa
\nn
P_{qq}^{<,\ve}(z) &=& - C_F (1-z), \SSP P_{qq}^{\dl,\ve} =
\fr{C_F}{2} \\
P_{qg}^{<,\ve}(z) &=& z^2 - z, \SSP P_{qg}^{\dl,\ve} = 0.
\eeqa
except that $P_{qg}^{<,\ve}$ depends on the convention used
for averaging over initial gluon states.
In the anticommuting $\gm_5$ scheme, one has
\beq
P_{qq}^n(z) = \Dl P^n_{qq}(z) = P_{q_+q_+}^n(z) -
\ub{P_{q_-q_+}^n(z)}_0
\eeq
(with the $\pm$'s denoting helicities) as a consequence of helicity
conservation (i.e.\ an anticommuting $\gm_5$).
This is not true in the HVBM scheme due to (\ref{e5}), which
violates helicity conservation of massless fermions (see
\cite{Vog} concerning the polarized split functions in the HVBM scheme).
We will show that this problem is overcome in the $\ovl{\rm MS}_\ve$
factorization scheme, which we shall now define.

Factorization of the mass singularities is equivalent to
expressing the bare parton distributions (and fragmentation
functions) in terms of the renormalized ones. In the
$\ovl{\rm MS}$ ($\ovl{\rm MS}_\ve$) scheme, this is done via
\beq
\label{e41}
[\Dl] f_{i/A}^0(x) = [\Dl] f_{i/A}^{\ovl{\rm MS}_{(\ve)}}(x,M_f^2)
+  \fr{c(\ve)}{\ve} \sum_j \fr{\al_s}{2\pi}\int_x^1\!\! \fr{dy}{y}
[\Dl] f_{j/A}^{\ovl{\rm MS}_{(\ve)}}(y,M_f^2) [\Dl] P_{ij}^{(n)}(x/y),
\eeq
where
\beq
\label{e42}
\fr{c(\ve)}{\ve}=\fr{1}{\ve} \lf(\fr{4\pi\mu^2}{M_f^2}\rt)^{\ve}
\fr{\Gm(1-\ve)}{\Gm(1-2\ve)}
= \lf(\fr{\mu^2}{M_f^2}\rt)^{\ve}
(\fr{1}{\ve} - \gm_{\rm E} + \ln 4\pi)
+ {\cal O}(\ve).
\eeq
The conventional factor $(\mu^2/M_f^2)^\ve$ is not necessary, but
it allows for a distiction between coupling renormalization and
mass factorization energy scales. We will take $M_f=\mu$ in our
calculations. In other processes though, this distinction might
be necessary in order to avoid large logarithms.

For the fragmentation functions ${\cal D}_{A/i}$, representing
the probability for quark $i$ to split into hadron $A$, the
corresponding renormalization is
\beq
\label{e43}
 [\Dl] {\cal D}_{A/i}^0(z)
= [\Dl] {\cal D}_{A/i}^{\ovl{\rm MS}_{(\ve)}}(z,M_f^2)
+  \fr{c(\ve)}{\ve} \sum_j \fr{\al_j}{2\pi}\int_z^1\!\! \fr{dy}{y}
[\Dl] {\cal D}_{A/j}^{\ovl{\rm MS}_{(\ve)}}(y,M_f^2)
[\Dl] P_{ji}^{(n)}(z/y),
\eeq
where $\al_j=\al_s$, unless $j$ (=$A$) = $\gm$, in which case
$\al_j = \al$.

It is clear that the $\ovl{\rm MS}_{\ve}$ scheme is just the
$\ovl{\rm MS}$ scheme, extended so as to subtract off the entire
$n$-dimensional split function. In DRED, $\ovl{\rm MS}_{\ve}$
is equivalent to $\ovl{\rm MS}$ since there is no
$\ve$-dimensional part of the split function.

We are now in a position to write down the results for
$[\Dl] d\ha{\sg}_{ab}/dM^2$ in a form valid for all $n$-dimensional
schemes. We start with the $q\B{q}$ subprocess. The Born term is
given by
\beq \label{e44}
[\Dl] \fr{d\ha{\sg}_{q\B{q}}}{dM^2} = [\Dl] \chi_B (\ve) \dl(1-w)
\eeq
with
\beq
\chi_B^{\rm DRED}(\ve) = -\Dl \chi_B^{\rm DRED}(\ve)
= \fr{\al^2}{N_c} \fr{e_q^2}{2^{-2\ve}} \fr{2\pi^{1+\ve}}{M^{4+2\ve}}
\mu^{4\ve} \fr{(2-\ve)}{(3-2\ve)(1-2\ve)} \fr{\Gm(1-\ve)}{\Gm(1-2\ve)}
\eeq
and
\beq
\chi_B^{\rm DREG}(\ve) = 2\fr{(1-\ve)^2}{(2-\ve)} \chi_B^{\rm DRED}(\ve).
\eeq
Here $N_c=3$ and $e_q$ is the quark fractional charge.
In the anticommuting $\gm_5$ scheme
\beq
\Dl \chi_B^{\rm AC}(\ve) = -\chi_B^{\rm DREG}(\ve).
\eeq
This is not true in the HVBM scheme though.
Of course, in the limit $\ve\RA 0$ helicity conservation is restored
in all schemes so long as there are no $1/\ve$ poles arising
from mass singularities. If there are such
$1/\ve$ poles, then one needs a scheme such as $\ovl{\rm MS}_\ve$,
as we will see.

The factorization counterterm in the $\ovl{\rm MS}$
($\ovl{\rm MS}_\ve$) scheme is
\beq
\label{e48}
[\Dl] \fr{d\ha{\sg}_{q\B{q}}^{\rm ct}}{dM^2}
= \fr{2}{\ve} [\Dl] \chi_B(\ve)  \fr{\al_s}{2\pi}
w^{1+\ve} [\Dl]  P_{qq}^{(n)}(w)
\lf(\fr{4\pi\mu^2}{M^2}\rt)^\ve \fr{\Gm(1-\ve)}{\Gm(1-2\ve)}
\lf(\fr{s}{M_f^2}\rt)^{\ve}.
\eeq
The gluonic bremsstrahlung contribution is
\beqa
\label{e49}
[\Dl] \fr{d\ha{\sg}_{q\B{q}}^{\rm Br}}{dM^2}
&=& [\Dl] \chi_B(\ve) \fr{\al_s}{2\pi} w^{1+\ve} C_F
\lf(\fr{4\pi\mu^2}{M^2}\rt)^\ve \fr{\Gm(1-\ve)}{\Gm(1-2\ve)} \\
\nn &\times&
\lf[\fr{2}{\ve^2}\dl(1-w)
-  \fr{2}{\ve}\fr{[\Dl] P_{qq}^<(w)}{C_F}
+  8\lf(\fr{\ln(1-w)}{1-w}\rt)_+
-4(1+w)\ln(1-w) - 2(1-w)
\rt]
.
\eeqa
The virtual contribution is
\beq
\label{e50}
[\Dl] \fr{d\ha{\sg}_{q\B{q}}^{\rm V}}{dM^2}
= [\Dl] \chi_B(\ve) \dl(1-w) C_F \fr{\al_s}{2\pi}
\lf(\fr{4\pi\mu^2}{M^2}\rt)^\ve \fr{\Gm(1-\ve)}{\Gm(1-2\ve)}
\lf[-\fr{2}{\ve^2}  - 7 + \fr{2\pi^2}{3} -\fr{2}{\ve}
\fr{P_{qq}^\dl}{C_F}  \rt]
.
\eeq
So, adding (\ref{e44}) and
 (\ref{e48}) -- (\ref{e50}) we obtain the total result
\beqa
\label{e51}
[\Dl] \fr{d\ha{\sg}_{q\B{q}}}{dM^2}
&=& [\Dl] \chi_B(0) \lf( \dl(1-w) +  C_F \fr{\al_s}{2\pi}w
\lf\{\lf( \fr{2\pi^2}{3}-7 \rt) \dl(1-w)  \rt.\rt.\\
\nn
&+& \lf.\lf. 8\lf(\fr{\ln(1-w)}{1-w}\rt)_+
+ 2 \fr{P_{qq}^4(w)}{C_F} \ln\fr{s}{M_f^2}
 - 4(1+w)\ln(1-w) - 2(1-w) + [\Dl] d_{q\B{q}} \rt\} \rt),
\eeqa
where
\beq
\label{e52}
[\Dl] d_{q\B{q}}^{\ovl{\rm MS}_\ve} = 0, \SSPP
[\Dl] d_{q\B{q}}^{\ovl{\rm MS}} = -\fr{2}{C_F}
([\Dl] P_{qq}^{<,\ve} +  P_{qq}^{\dl,\ve} \dl(1-w)).
\eeq
We see that the ${\cal O}(\ve)$ scheme dependence of the Born term
cancels with the $1/\ve$ ($1/\ve^2$) divergences multiplying it.
Also, we see that all the $n$-dimensional regularization schemes give
the same answer in the $\ovl{\rm MS}_\ve$ scheme and it corresponds
to the DRED $\ovl{\rm MS}$ answer.

We now consider $[\Dl] d\ha{\sg}_{qg}/dM^2$. There is no
${\cal O}(1)$ term (in $\al_s$). At ${\cal O}(\al_s)$ there is a
factorization counterterm contribution which is given in the
$\ovl{\rm MS}$ ($\ovl{\rm MS}_\ve$) scheme, by
\beq
\label{e53}
[\Dl] \fr{d\ha{\sg}_{qg}^{\rm ct}}{dM^2}
= \fr{1}{\ve} [\Dl] \chi_B(\ve)  \fr{\al_s}{2\pi}
w^{1+\ve} [\Dl] P_{qg}^{(n)}(w)
\lf(\fr{4\pi\mu^2}{M^2}\rt)^\ve \fr{\Gm(1-\ve)}{\Gm(1-2\ve)}
\lf(\fr{s}{M_f^2}\rt)^{\ve}.
\eeq
The bremsstrahlung contribution is
\beqa
\label{e54}
\nn
[\Dl] \fr{d\ha{\sg}_{qg}^{\rm Br}}{dM^2}
&=&  [\Dl] \chi_B(\ve) \fr{\al_s}{2\pi} w^{1+\ve}
\lf(\fr{4\pi\mu^2}{M^2}\rt)^\ve \fr{\Gm(1-\ve)}{\Gm(1-2\ve)} \\
&\times&
\lf[-\fr{1}{\ve}[\Dl] P_{qg}^<(w)
+  2 [\Dl] P_{qg}^4(w)\ln(1-w) + \fr{(1-w)^2}{4} + w(1-w)
\rt]
.
\eeqa
Adding (\ref{e53}) and (\ref{e54}), we obtain the total result
\beq
\label{e435}
[\Dl] \fr{d\ha{\sg}_{qg}}{dM^2}
= [\Dl] \chi_B(0)  \fr{\al_s}{2\pi} w
\lf\{
[\Dl] P_{qg}^4 (w)\lf[\ln\fr{s}{M_f^2} +2\ln(1-w)\rt]
+\fr{(1-w)}{4}(1+3w) + [\Dl] d_{qg}
\rt\},
\eeq
where
\beq
[\Dl] d_{qg}^{\ovl{\rm MS}_\ve} =0, \SSP
[\Dl] d_{qg}^{\ovl{\rm MS}} = -[\Dl] P_{qg}^{<,\ve}(w).
\eeq

In both the $q\B{q}$ and $qg$ cases, we verify that the unpolarized
DREG $\ovl{\rm MS}$ result agrees exactly with that previously
determined (see for example \cite{Hamb}). Since
$P_{qq}^n \neq \Dl P_{qq}^n$ in the HVBM scheme, we see from
(\ref{e51}), (\ref{e52}) that $d\ha{\sg}_{q\B{q}}/dM^2
\neq - \Dl d\ha{\sg}_{q\B{q}}/dM^2$ in the HVBM scheme if
one uses $\ovl{\rm MS}$. But this is a physical requirement. Hence,
if one uses HVBM regularization, then it is necessary to use a
subtraction scheme like $\ovl{\rm MS}_\ve$ or one which subtracts {\em at
least} the helicity non-conserving part,
$\Dl P_{qq}^{<,\ve} -  P_{qq}^{<,\ve}$ as well
(see also \cite{Vog}) in the polarized case.
 Of course, it makes more sense to subtract
the entire $\Dl P_{qq}^n$ since this leads to regularization
scheme independent results.

Now let us consider the Drell-Yan process with transversely
polarized hadrons ({\em transverse Drell-Yan}). The general
process is
\beq
\label{e57}
A(P_1,S_1) + B(P_2,\pm S_2) \RA l^-(p_3) + l^+(p_4) + X
\eeq
where the $S_i$ are reference spin vectors satisfying
\beq
S_i^2 = -1, \SSPP S_i\C P_j =0 \SSP i,j=1,2
\eeq
implying that $S_1$ and $S_2$ lie in the plane transverse to
the beam axis. Now, letting $\ua$ denote polarization in the
direction of the spin axis and letting $\da$ denote polarization
opposite to the spin axis, we may define the transversely polarized
cross section as
\beq
\label{e59}
\Dl_{\rm T} \sg \equiv \fr{1}{2} [\sg(\ua,\ua) - \sg(\ua,\da)],
\eeq
in the notation $\sg(S_1,\pm S_2)$.

The general $2\RA 2$ [$2\RA 3$]
subprocess contributing to (\ref{e57}) is
\beq
q(p_1,s_1) + \B{q}(p_2,\pm s_2) \RA \gm^*(q) + [g(k)]
\RA l^-(p_3) + l^+(p_4) + [g(k)]
\eeq
with
\beq
s_1 = S_1, \SSP s_2 =  S_2
\eeq
and $\Dl_{\rm T} \ha{\sg}$ defined analogously to
(\ref{e59}). There is no $qg$ subprocess since gluons cannot be
transversely polarized. We may define the transversity
distribution
\beq
\Dl_{\rm T} f_{q/A}(x,M_f^2)
\equiv \Dl_{\rm T} F_{q/A} (x,M_f^2)/x =
f_{q_\ua/A_\ua}(x,M_f^2)
- f_{q_\da/A_\ua}(x,M_f^2),
\eeq
which is often denoted as $h_1^q(x,M_f^2)$.

Let $\phi_3$ denote the azimuthal angle of $p_3$
about the beam axis (with respect to some reference axis)
and let $\ha{\theta}_3$ denote the angle
between $p_1$ and $p_3$ in the $p_1$, $p_2$ center-of-momentum frame.
Then, the quantity
we are interested to calculate is $\Dl_{\rm T}d\ha{\sg}/dM^2d\phi_3$.
This was done in \cite{PLB} using DRED, where
$\Dl_{\rm T}d\ha{\sg}/dM^2d\ha{\Omega}_3$ was first determined, with
$\ha{\Omega}_3$ representing the solid angle of $p_3$ in the
c.m.\ of $p_1$, $p_2$. Then $\Dl_{\rm T} d\ha{\sg}/dM^2d\phi_3$ was
obtained via
\beq
 \fr{\Dl_{\rm T}d\ha{\sg}}{dM^2d\phi_3} =
\int_{-1}^1 d(\cos\ha{\theta}_3)
 \fr{\Dl_{\rm T}d\ha{\sg}}{dM^2d\ha{\Omega}_3}.
\eeq
Of course, if one was only interested
in $\Dl_{\rm T} d\ha{\sg}/dM^2d\phi_3$, then one could integrate
over $\ha{\theta}_3$ separately for the Born term, loops,
bremsstrahlung and factorization counterterm, then add all
the different parts to get a finite result for
$\Dl_{\rm T} d\ha{\sg}/dM^2d\phi_3$.
Either way, the expression for $\Dl_{\rm T}\sg_{AB}/dM^2d\phi_3$
is obtained from (\ref{e20}) by replacing $[\Dl]\RA \Dl_{\rm T}$
and differentiating with respect to $\phi_3$.

{}From the form of the unpolarized and longitudinally polarized
results, it is straightforward just to take the final
DRED result of \cite{PLB} and put it in a form valid for all
regularization schemes. The result is
(with $\phi_1$, $\phi_2$ the azimuthal angles of $s_1$, $s_2$)
\beqa
\label{e64}
\fr{\Dl_{\rm T} d\ha{\sg}}{dM^2d\phi_3}
&=& \Dl_{\rm T} \chi_B \lf( \dl(1-w) + \fr{\al_s}{2\pi} C_F w
\lf\{ \lf[\fr{2\pi^2}{3} -7 \rt]\dl(1-w) + 8 \lf( \fr{\ln(1-w)}
{1-w} \rt)_+  \rt. \rt. \\
\nn &+& \lf.\lf.
 2 \fr{\Dl_{\rm T}P^4_{qq}(w)}{C_F} \ln \fr{s}{M_f^2}
- 8\ln(1-w) - 6w\fr{\ln^2w}{1-w} + 4 (1-w) + \Dl_{\rm T} d
\rt\} \rt),
\eeqa
with
\beq
\Dl_{\rm T} \chi_B = \cos(\phi_1+\phi_2-2\phi_3) \fr{\al^2}{3N_c}
\fr{e_q^2}{M^4},
\eeq
and
\beq
\Dl_{\rm T} d_{\ovl{\rm MS}_\ve} = 0,
\SSPP \Dl_{\rm T} d_{\ovl{\rm MS}} = - \fr{2}{C_F}
(\Dl_{\rm T} P_{qq}^{<,\ve} +  P_{qq}^{\dl, \ve}
\dl(1-w)).
\eeq
The transversity split function $\Dl_{\rm T} P^n_{qq}$ is obtained
via (\ref{e34}) with $[\Dl] \RA \Dl_{\rm T}$.
In 4 dimensions, $\Dl_{\rm T} P^4_{qq}$ is given by \cite{Art}
\beq
\Dl_{\rm T} P^4_{qq}(z) = C_F \lf[ \fr{2}{(1-z)_+} -2
+ \fr{3}{2} \dl(1-z) \rt].
\eeq
In the anticommuting $\gm_5$ scheme, it is straightforward
to obtain
\beq
\Dl_{\rm T} P_{qq}^{<,\ve}(z) = -C_F(1-z).
\eeq

The transversity
renormalizations corresponding to (\ref{e41}), (\ref{e43})
are obtained by replacing $[\Dl] \RA \Dl_{\rm T}$. We verified
explicitly that the form of (\ref{e41}) does indeed hold
for the transverse case.

\vglue 1cm
\begin{center}\begin{large}\begin{bf}
IV. CONNECTION BETWEEN $n$-DIMENSIONAL SCHEMES
\end{bf}\end{large}\end{center}
\vglue .3cm

Here we will show how to convert results of one $n$-dimensional
scheme to those of another in a straightforward manner. We do
this by examining the origin of the scheme dependent parts. Strictly
speaking, this only applies to processes not requiring
coupling constant renormalization, since other differences may
arise from the UV sector. On the other hand, it has been shown
\cite{Jones,ACMP}
that the UV sectors of DREG and DRED in QCD can be related
via a finite ${\cal O}(\al_s^2)$
renormalization of the coupling.
Namely,
\beq
\al_s^{\rm DRED} = \al_s^{\ovl{\rm MS}}(1+\fr{\al_s^{\ovl{\rm MS}}}{2\pi}
\fr{N_c}{6})
\eeq
(see for example \cite{Kun}). In other words, one may go from one
scheme to the other by simply expressing the coupling of one
scheme in terms of that in the other.
We will assume this has
been done so that the only differences may arise from the IR
sector. Then all the following argumentation can be
seen to apply to all one-loop QCD processes.

As an example, we will consider the $q\B{q}$ subprocess in
unpolarized Drell-Yan, to show the origin of the scheme
dependences. Then we will show that the same argumentation
holds for all one-loop processes.

In order to extract the scheme dependent parts, we need only consider
terms which give rise to $1/\ve$ poles. This is because the
scheme dependences come from $\fr{1}{\ve}\C {\cal O}(\ve)$ terms,
where the ${\cal O}(\ve)$ terms are in general scheme dependent.
We therefore consider the contribution to $d\ha{\sg}_{q\B{q}}/dM^2$
when $k$ is collinear with one of  the initial partons, say
$p_1$. From (\ref{e34}) we see that
\beq
|M|^2_{2\RA 3} \sim \fr{P_{qq}^<(w,\ve)}{p_1\C k}
|M|^2_{\rm Born}, \SSPP w<1
\eeq
with
\beq
k \approx (1-w)p_1.
\eeq
After phase space integrations, this will yield a contribution
to $d\ha{\sg}_{q\B{q}}/dM^2$,
\beqa
\fr{d\ha{\sg}_{q\B{q}}^{\rm coll}}{dM^2}
&\sim& \fr{1}{\ve} \fr{\chi_B(\ve)}{(1-w)^{1+2\ve}}
[(1-w)P_{qq}^< (w,\ve)] \\
\nn
&=& \fr{\chi_B(\ve)}{\ve} \lf( - \fr{1}{2\ve} \dl(1-w)
+ \fr{1}{(1-w)_+} - 2\ve\lf( \fr{\ln(1-w)}{1-w}\rt)_+\rt)
[(1-w)P_{qq}^<(w,\ve)].
\eeqa
Hence (noting that $P_{qq}^{<,\ve}(w) \sim 1-w$) we get
a scheme dependent part from the bremsstrahlung,
\beq
\fr{d\ha{\sg}_{\rm Br}^{\rm SD}}{dM^2}
\sim \chi_B(0) P_{qq}^{<,\ve}(w)
\eeq
as well as an equal term arising from $k\approx (1-w)p_2$. The soft
divergent terms $\sim (\fr{1}{\ve} \dl(1-w)) \C \fr{1}{\ve}$
cancel exactly with the loops, having the same overall factor,
and hence do not lead to scheme dependences.

As well, there is a scheme dependent term coming from the loops.
By definition of $P_{qq}^\dl$, there must be a term proportional
to
\beq
\fr{d\ha{\sg}_{\rm V}}{dM^2}
\sim \chi_B(\ve)
 \fr{P_{qq}^\dl}{\ve} \dl(1-w).
\eeq
This will lead to a scheme dependent part
\beq
\fr{d\ha{\sg}_{\rm V}^{\rm SD}}{dM^2}
\sim \chi_B(0) P_{qq}^{\dl,\ve} \dl(1-w).
\eeq

Hence, the entire scheme dependence can be traced back to
the process-independent $n$-dimensional split functions
(their $\ve$-dimensional part). We also see explicitly why
the $\ovl{\rm MS}_\ve$ factorization scheme will lead to
regularization scheme independent results; all the regularization
scheme dependent parts are subtracted. Of course, if one has
longitudinal or transverse polarization, all the above holds
with
\beq
\chi_B(\ve) \RA \Dl_{[\rm T]} \chi_B(\ve), \SSPP
P_{qq} (w,\ve) \RA \Dl_{[\rm T]} P_{qq}(w,\ve).
\eeq
For conciseness, we will drop the $\Dl_{[\rm T]}$'s with the understanding
that the same argumentation holds for the polarized cases.

It is now clear how to convert results calculated in DRED to
the corresponding  DREG $\ovl{\rm MS}$ results, if desired.
One simply replaces (defining $P_{ij}^\ve(z) \equiv
P_{ij}^{<,\ve}(z) + P_{ij}^{\dl,\ve} \dl(1-z)$)
\beq
\label{e73}
P_{ij}^4(z) \RA P_{ij}^4(z) - \ve P_{ij}^{\ve}(z)
\eeq
in the factorization counterterm.

Or, to convert from DREG to DRED (i.e.\ $\ovl{\rm MS}_\ve$),
simply replace
\beq
\label{e74}
P_{ij}^4(z) \RA P_{ij}^n(z,\ve) = P_{ij}^4(z) + \ve P_{ij}^\ve(z)
\eeq
in the factorization counterterm. Hence one can go from any
$n$-dimensional scheme to any another using (\ref{e73})
and (\ref{e74}). From the previous argumentation, it is clear
that polarization poses no difficulties in this approach.
This procedure is equivalent to expressing the parton
distributions and fragmentation functions of one scheme in terms
of those in the other, as will be shown in the next section.

Up until now, we have used Drell-Yan as an example. We now show
that the argumentation here applies to all QCD processes at one-loop
once the UV sectors have been made to agree (if coupling
constant renormalization is required). The renormalization of
the parton distributions is process independent, since
the same collinear configurations occur in all processes. This
is because they are related to hadronic emissions (or
fragmentation into hadrons) which occur in a process independent
manner and can be constructed from a universal set of subdiagrams
containing the configurations having collinear divergences
with respect to the particular parton.
The only difference then, from one process to the next, is the
soft singularity structure. But the soft singularities cancel
via the Bloch-Nordseick mechanism \cite{BN} (or KLN theorem
\cite{KLN}) such that the singular
cross section for soft emissions is proportional to the Born
term with (minus) the same overall factor as that coming from
the loops. Hence, the scheme dependences of ${\cal O}(\ve)$ which
which multiply the $1/\ve$ singular terms, cancel exactly between
the loops and the soft bremsstrahlung contributions.

Thus, the only scheme dependence may come from the non-cancelling
mass singularities, whose structure is process independent.
Hence, all the argumentation here applies to all QCD processes
at one-loop; extension to higher orders should be analogous.
Also, the same conclusions concerning the regularization
scheme independence in the $\ovl{\rm MS}_\ve$ scheme apply
to all one-loop QCD processes.

\newpage

\vglue 1cm
\begin{center}\begin{large}\begin{bf}
V. DRED PARTON DISTRIBUTIONS AND FRAGMENTATION FUNCTIONS
\end{bf}\end{large}\end{center}
\vglue .3cm

It is straightforward to show how to relate the DRED parton
distributions and fragmentation functions to those in DREG
($\ovl{\rm MS}$). This is useful if one wishes to work strictly
in DRED, but make use of the abundant sets of DREG parton
distributions and fragmentation functions. First, we will
consider the parton distributions, again dropping the
$\Dl_{[\rm T]}$'s for conciseness.

We will make use of the fact that DRED is equivalent to the
$\ovl{\rm MS}_\ve$ scheme of DREG.
Noting that the bare parton distribution, $f_{i/A}^0(x)$,
on the LHS of (\ref{e41}) is factorization scheme independent, we
obtain
\beq
f_{i/A}^{\rm DRED}(x) =  f_{i/A}^{\ovl{\rm MS}_\ve}(x)
= f_{i/A}^{\ovl{\rm MS}}(x)
+  \sum_j \fr{\al_s}{2\pi}\int_x^1\!\! \fr{dy}{y} \fr{c(\ve)}{\ve}
(f_{j/A}^{\ovl{\rm MS}}(y) P_{ij}^4(x/y)
- f_{j/A}^{\ovl{\rm MS}_\ve}(y)  P_{ij}^n(x/y) )
\eeq
or
\beq \label{fia}
f_{i/A}^{\rm DRED}(x)
=  f_{i/A}^{\ovl{\rm MS}_\ve}(x)
= f_{i/A}^{\ovl{\rm MS}}(x)
-  \sum_j \fr{\al_s}{2\pi}\int_x^1\!\! \fr{dy}{y}
f_{j/A}^{\ovl{\rm MS}}(y) P_{ij}^\ve(x/y)
+ {\cal O} (\al_s^2).
\eeq
{}From (\ref{e43}), we may immediately write for the fragmentation
functions
\beq \label{dia}
{\cal D}_{A/i}^{\rm DRED}(z)
= {\cal D}_{A/i}^{\ovl{\rm MS}_\ve}(z)
= {\cal D}_{A/i}^{\ovl{\rm MS}}(z)
-  \sum_j \fr{\al_j}{2\pi}\int_z^1\!\! \fr{dy}{y}
{\cal D}_{A/j}^{\ovl{\rm MS}}(y) P_{ji}^\ve(z/y)
+ {\cal O} (\al_s^2).
\eeq
Comparing with (\ref{e73}), (\ref{e74}) we see explicitly that going from
DRED to DREG $\ovl{\rm MS}$ simply amounts to expressing the
DRED (or $\ovl{\rm MS}_\ve$) parton distributions and fragmentation
functions in terms of the DREG $\ovl{\rm MS}$ ones, and vice-versa
from series inversion.

\vglue 1cm
\begin{center}\begin{large}\begin{bf}
VI. NUMERICAL RESULTS
\end{bf}\end{large}\end{center}
\vglue .3cm

Here we present asymmetries and cross sections for the Drell-Yan
process in $p$-$p$ collisions at energies relevant to RHIC.
In general,
we use the two-loop $\ovl{\rm MS}$ expression for $\al_s(\mu^2)$,
with four flavors and $\Lambda=0.2\,\,\GeV$, except in the transversely
polarized cross sections where we use the one-loop expression in
order to be consistent with \cite{PLB}.
Also, we take $\mu^2=M_f^2=M^2$.
For the unpolarized cross sections, we use the DREG subprocess cross
section convoluted with the unpolarized parton distributions of
\cite{MT} (Set S-$\ovl{\rm MS}$). For the longitudinally polarized
case, we use the $\ovl{\rm MS}_\ve$ (or DRED) subprocess cross
sections, since they are physically consistent (and
regularization scheme independent), convoluted with the longitudinally
polarized parton distributions of \cite{AS} (Set 1, $SU(3)$ symmetric sea)
which fit well the recent DIS data \cite{DIS} except at very low $x$ not
covered for the kinematics considered here.

For the transversely polarized subprocess cross sections, we again
use the $\ovl{\rm MS}_\ve$ result.
For the transversity distributions, we choose
for the valence distributions (at $Q_0^2 = 4\,\,\GeV^2$)
\beqa
\Dl_{\rm T} F_{u_v/p}(x,Q_0^2) &=& 2.1\, x^{.8} (1-x)^{2.4} \\
\Dl_{\rm T} F_{d_v/p}(x,Q_0^2) &=& -.76\, x^{.8} (1-x)^{3.4}
\eeqa
and for the sea-quark distributions
\beq
\Dl_{\rm T} F_{\B{q}/p}(x,Q_0^2) = -.12 \, x^{.1} (1-x)^{9.5},
\SSP q=u,d,s
\eeq
(one half the value used in \cite{PLB,Thesis}). These satisfy the
upper bound proposed by Soffer \cite{Sof}. As well, the valence
 distributions
are consistent with bag model predictions. The small- and large-$x$
behaviour is consistent with the longitudinal and unpolarized
cases. There are no other definite constraints we may impose.
$Q^2$-dependent parameterizations for the longitudinal and
transversity distributions are given in \cite{Thesis}.
We use one-loop evolutions, as two-loop polarized split functions
do not yet exist.

Define the asymmetries
\beq
A_T = \fr{\Dl_{\rm T} d\sg/dM^2d\phi_3}{d\sg/dM^2d\phi_3}
\eeq
and
\beq
A_L = \fr{\Dl d\sg/dM^2}{d\sg/dM^2},
\eeq
noting that
\beq
\fr{d\sg}{dM^2d\phi_3} = \fr{1}{2\pi} \fr{d\sg}{dM^2}.
\eeq

Fig.\ 1(a) presents $A_T$ for $\sqrt{S} = 100 \,\, \GeV$, in the
range $0.05 \leq \sqrt{\tau} \leq 0.5$. The largest value for
$-A_T$ is 8\%. We notice that for the $q\B{q}$ subprocess,
$A_T$ is reasonably perturbatively stable, unlike the cross sections
which increase by 50 -- 100 \% under HOC \cite{PLB}. Inclusion of the $qg$
subprocess makes the asymmetry somewhat more negative since the
$qg$ subprocess contributes negatively as noticed in \cite{AEM,Kub}.

Fig.\ 1(b) presents the corresponding polarized cross section,
$-\Dl_{\rm T} d\sg/dM^2 d\phi_3$. The sharp
dropoff with increasing $\sqrt{\tau}$ is a combined result of
the softness of $\Dl_{\rm T} F_{\B{q}/p}$, the $1/M^4$ behaviour
of the cross section and the decreasing integration region
with increasing $\sqrt{\tau}$.

Fig.\ 2 presents the corresponding quantities for
$\sqrt{S} = 200 \,\, \GeV$ and $0.05 \leq \sqrt{\tau}
\leq 0.25$ (away from the $Z$-exchange region).
Similar features hold, except the cross sections
are somewhat smaller due to the $1/M^4$ suppression (which
amounts to $2/M^3$ in $\Dl_{\rm T} d\sg/dMd\phi_3$).

In Figures 3 and 4 we present the corresponding plots for $A_L$
and $-\Dl d\sg/dM^2$. The largest value for $-A_L$ is 16\%. As expected
from helicity conservation, the $q\B{q}$ subprocess exhibits
great perturbative stability. Inclusion of the $qg$ subprocess
however, upsets this stability since it contributes with sign opposite
to that of the $q\B{q}$ subprocess (in both the polarized and
unpolarized cases)
and is relatively large in the polarized case.
Hence the net
asymmetry becomes somewhat smaller in magnitude.
This wouldn't be a problem in $p$-$\B{p}$ collisions, where one
is probing valence-valence distributions. For $p$-$p$ collisions,
the smallness of $\Dl F_{\B{q}}$ makes the $qg$ subprocess more
significant. A smaller polarized gluon distribution and/or a
larger polarized sea-quark distribution would reduce this effect.
Still, for the larger $\sqrt{\tau}$, measurable
asymmetries are obtained.

For a discussion of the perturbative stablilty
of the longitudinal asymmetry in $p$-$p$ direct photon production, see
\cite{VogN,Thesis}. There it was noted that the asymmetry is perturbatively
stable if the polarized gluon distribution is sufficiently large that
the $qg$ subprocess dominates.
This contrasts with $p$-$p$ Drell-Yan, where a large polarized gluon
distribution destabilizes the asymmetry.

\vglue 1cm
\begin{center}\begin{large}\begin{bf}
VII. CONCLUSIONS
\end{bf}\end{large}\end{center}
\vglue .3cm

We have presented complete NLO
analytical results for the Drell-Yan process with unpolarized,
longitudinally polarized and transversely polarized hadrons. These
results are in a form valid for all $n$-dimensional schemes. It
was shown how one can easily convert from results obtained in
one scheme to those of another, regardless of the polarization,
for one-loop QCD processes.
This procedure simply amounts to expressing the parton distributions
and fragmentation functions in one scheme in terms of those
in the other.
As well, the origin of the scheme dependences was elucidated.
A mass factorization scheme, which we call $\ovl{\rm MS}_\ve$,
was introduced. It was shown that in this factorization scheme,
the final results are regularization scheme independent and coincide
with those of DRED $\ovl{\rm MS}$.
A simple method for converting parton distributions and fragmentation
functions from DREG to DRED was given.

For $p$-$p$ collisions at energies relevant to RHIC, asymmetries
and cross sections for transversely and longitudinally polarized
collisions were presented. For the transverse case, the asymmetries
reached -8\% and exhibited reasonable
perturbative stablity. For the longitudinal
case, the asymmetries reached -16\% and the $q\B{q}$ subprocess
exhibited great perturbative stability. Inclusion of the $qg$
subprocess somewhat lessened the longitudinal asymmetries however.
Still, $p$-$p$ collisions serve as the best probe of the polarized
antiquark distributions in the proton, and they may be extracted
with sufficient experimental statistics.

\vglue 1cm
\begin{center}\begin{large}\begin{bf}
ACKNOWLEDGEMENTS
\end{bf}\end{large}\end{center}
\vglue .3cm
The author thanks A.P.\ Contogouris and Z.\ Merebashvili for
enlightening discussions as well as G.\ Bunce for useful
conversations.
This work was also supported by the
Natural Sciences and Engineering Research Council of Canada and
by the Qu\'{e}bec Department of Education.

\vglue 1cm
\begin{center}\begin{large}\begin{bf}
REFERENCES
\end{bf}\end{large}\end{center}
\vglue .3cm

\vglue 1cm
\begin{center}\begin{large}\begin{bf}
FIGURE CAPTIONS
\end{bf}\end{large}\end{center}
\vglue .3cm

\begin{thecaptions}{9}
\item{} {\bf (a)} Transverse asymmetry, $A_T$, in leading order and in
next-to-leading order; {\bf (b)} corresponding next-to-leading
order polarized cross section versus $\sqrt{\tau}$ at
$\sqrt{S}=100\,\,\GeV$.
\item{} {\bf (a)} Transverse asymmetry, $A_T$, in leading order and in
next-to-leading order; {\bf (b)} corresponding next-to-leading
order polarized cross section versus $\sqrt{\tau}$ at
$\sqrt{S}=200\,\,\GeV$.
\item{} {\bf (a)} Longitudinal asymmetry, $A_L$, in leading order and in
next-to-leading order; {\bf (b)} corresponding next-to-leading
order polarized cross section versus $\sqrt{\tau}$ at
$\sqrt{S}=100\,\,\GeV$.
\item{} {\bf (a)} Longitudinal asymmetry, $A_L$, in leading order and in
next-to-leading order; {\bf (b)} corresponding next-to-leading
order polarized cross section versus $\sqrt{\tau}$ at
$\sqrt{S}=200\,\,\GeV$.
\end{thecaptions}

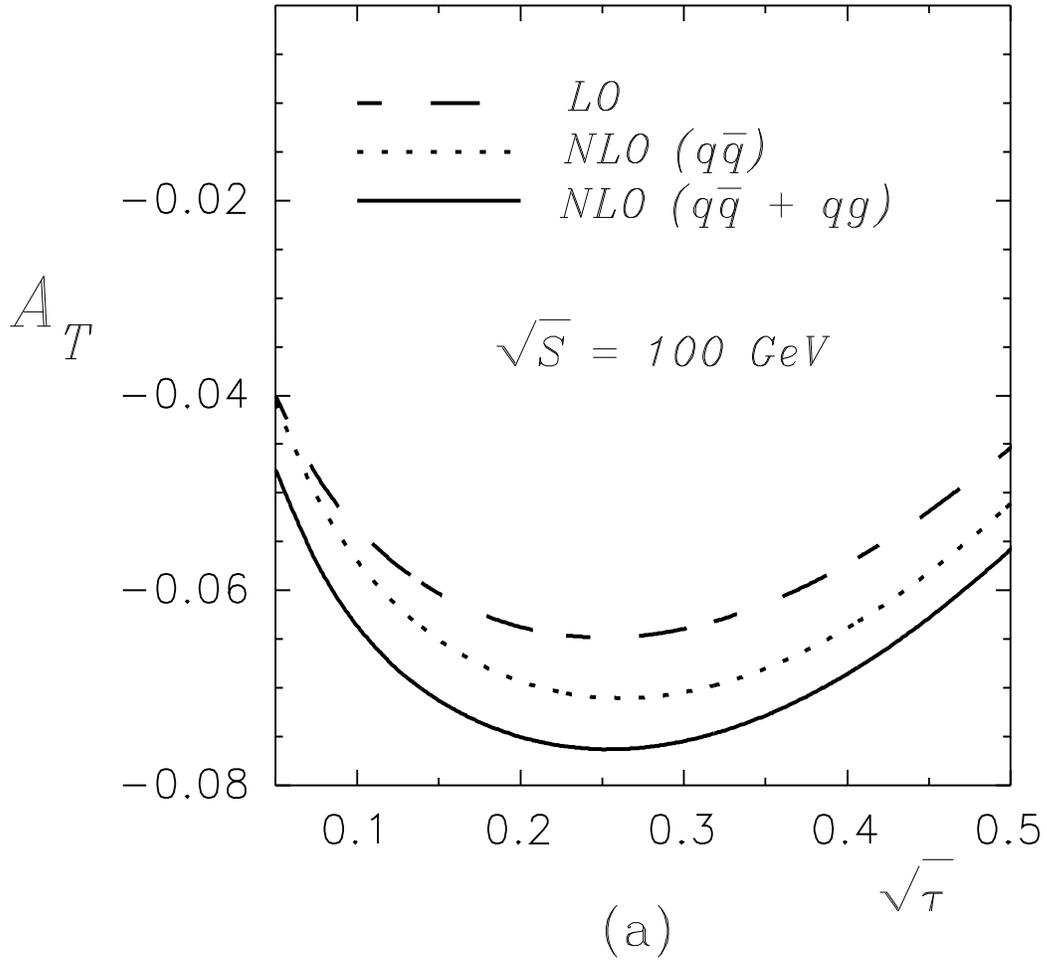
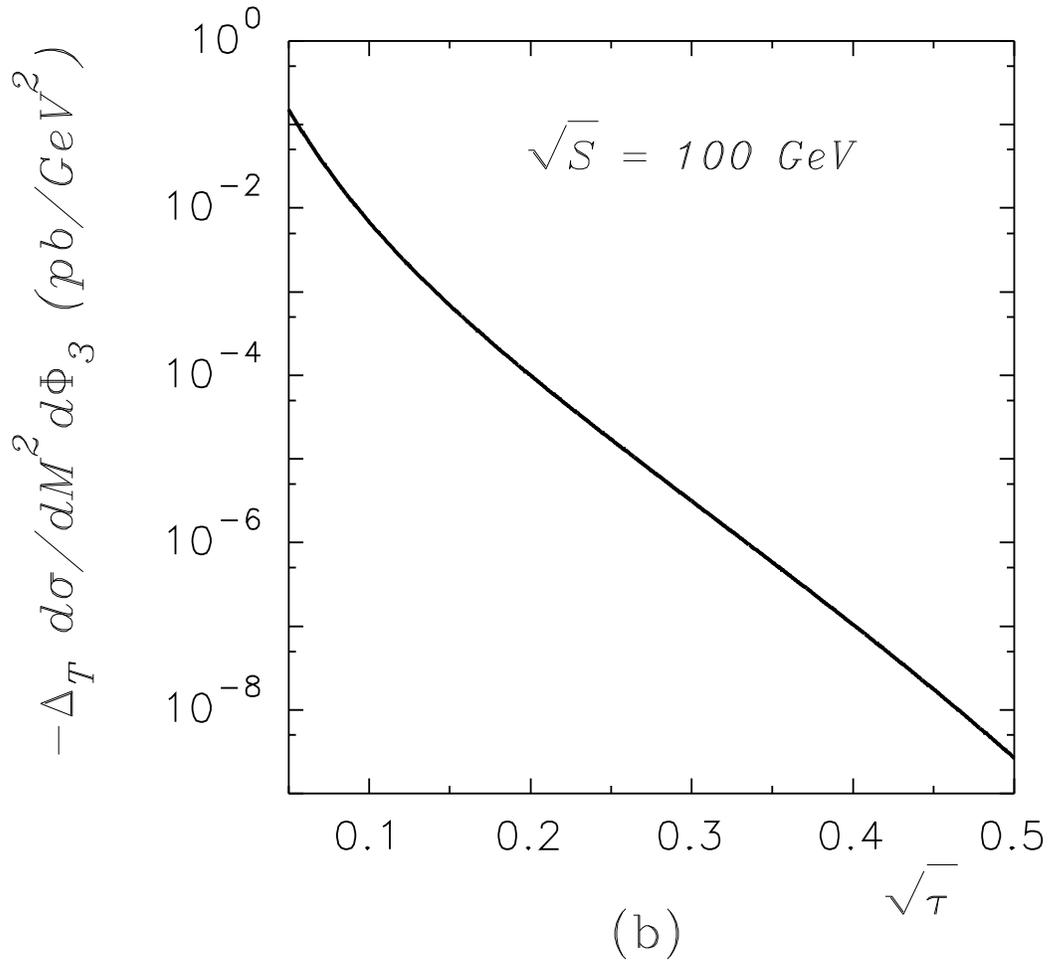
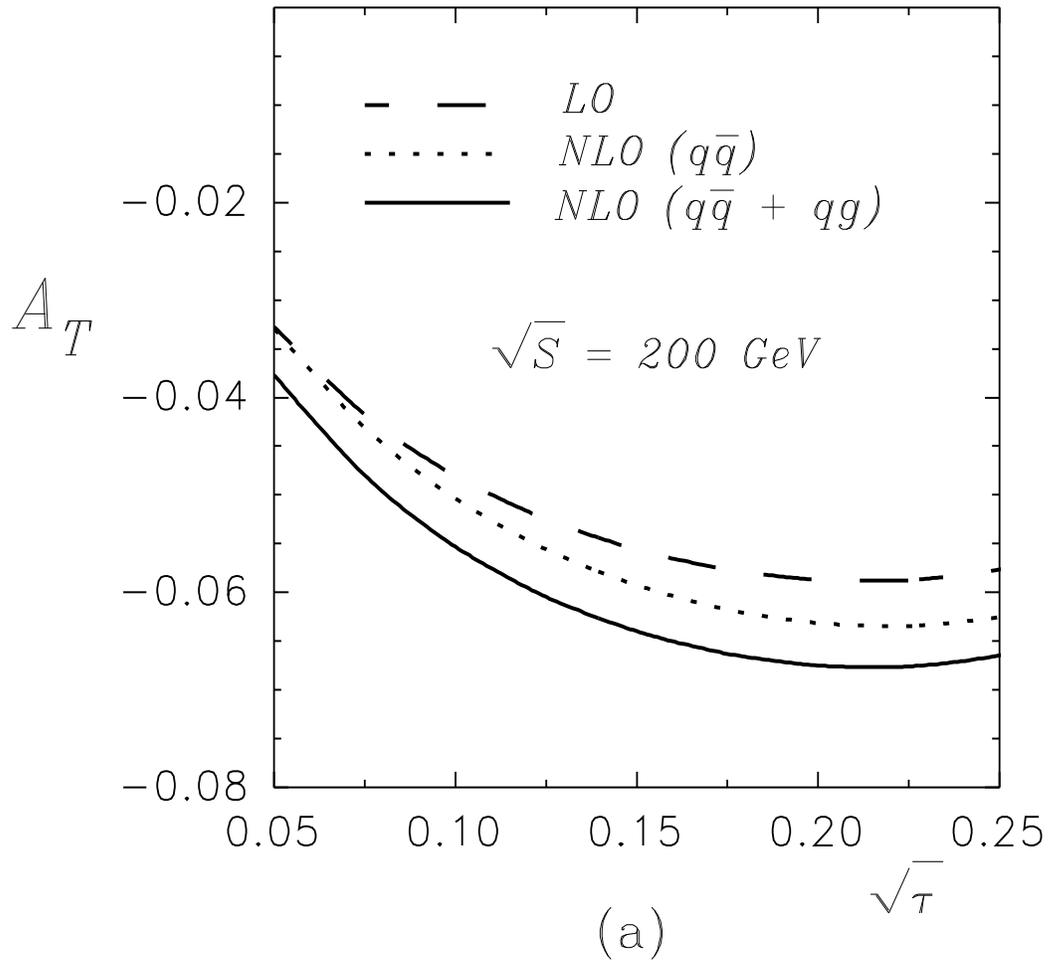
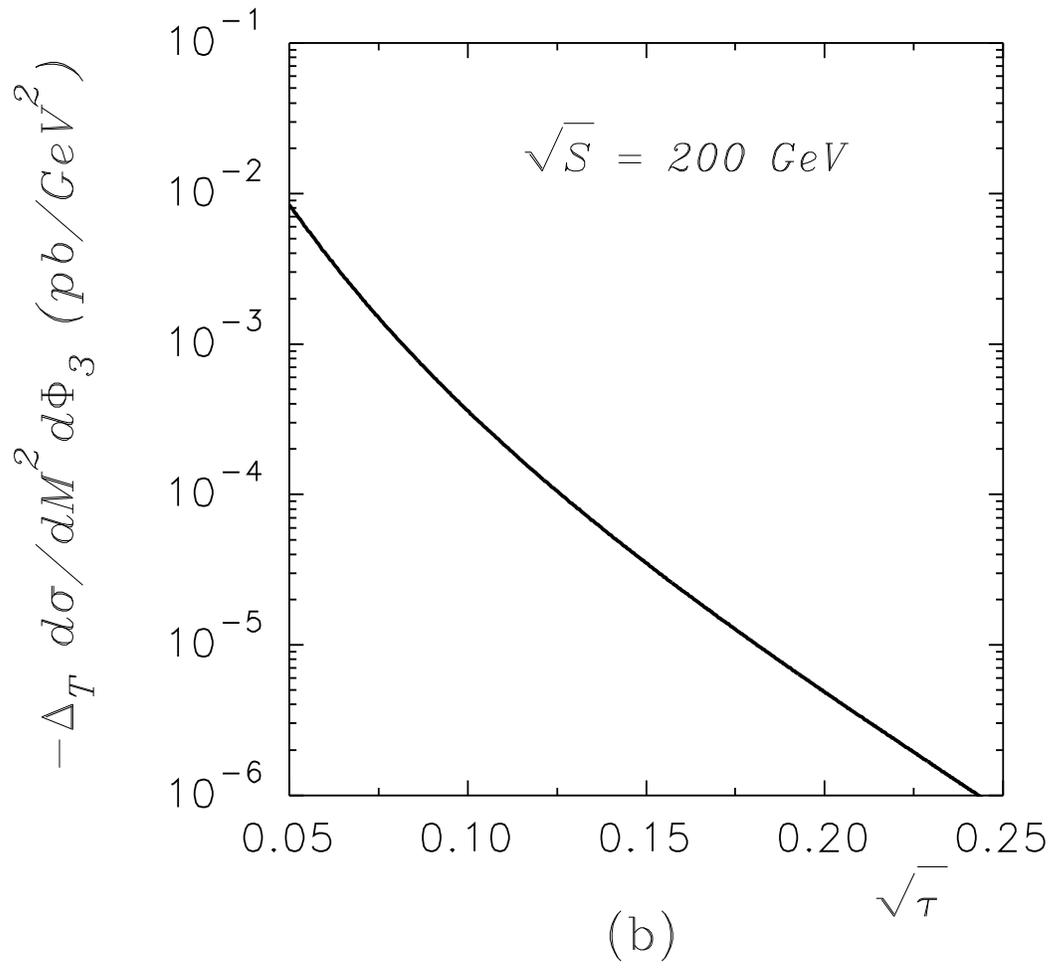
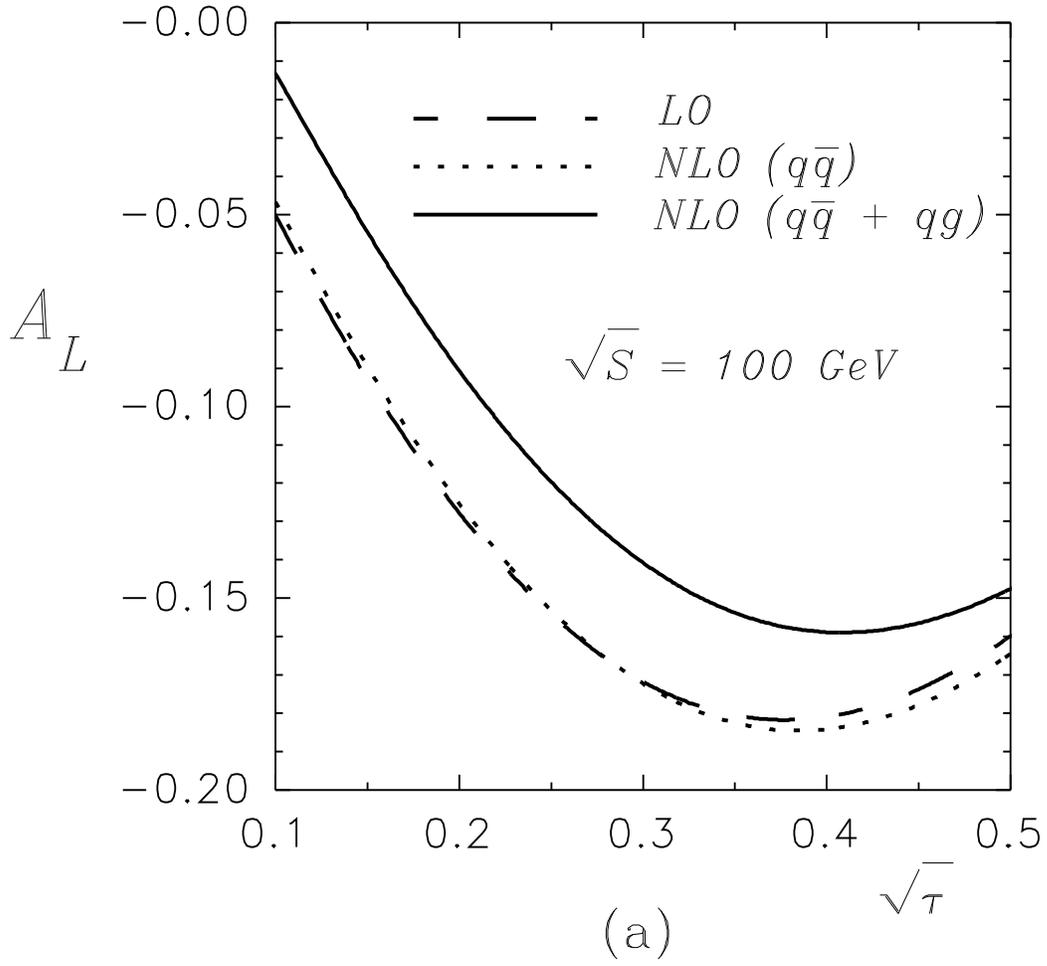
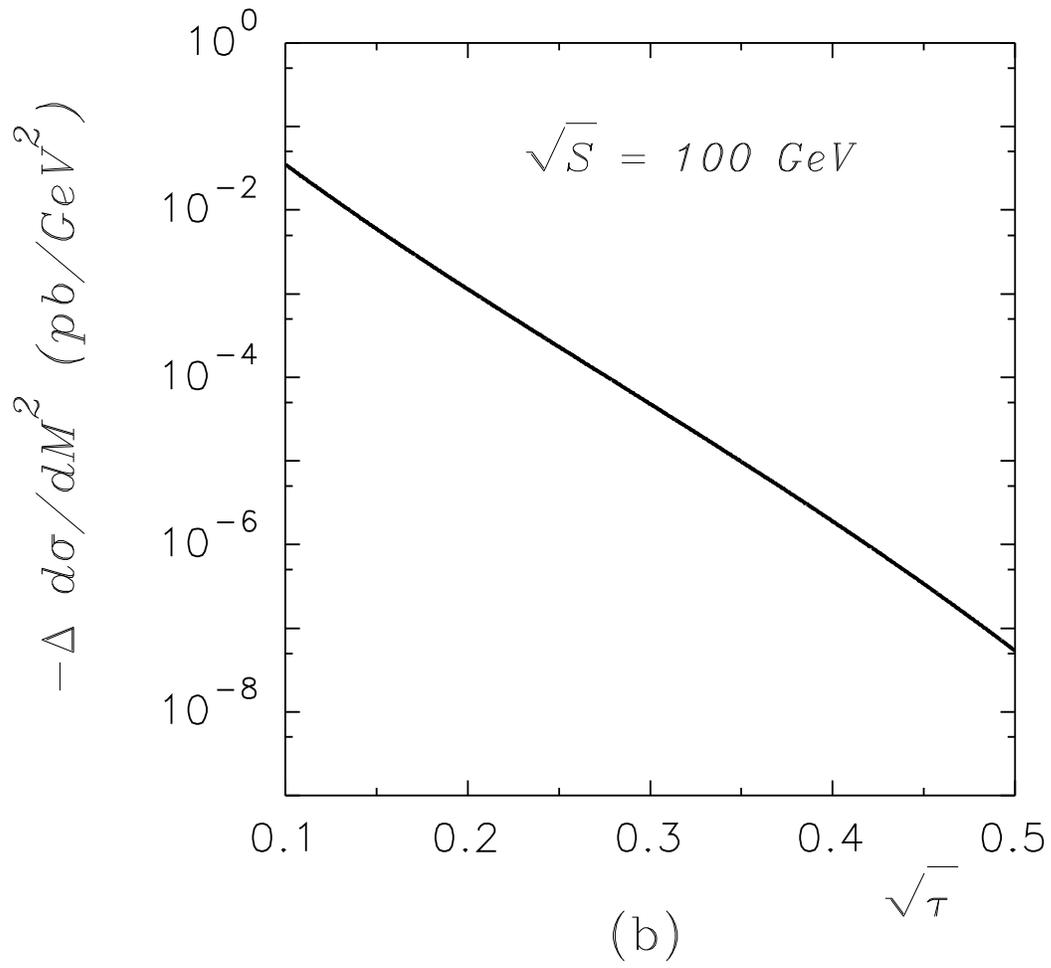
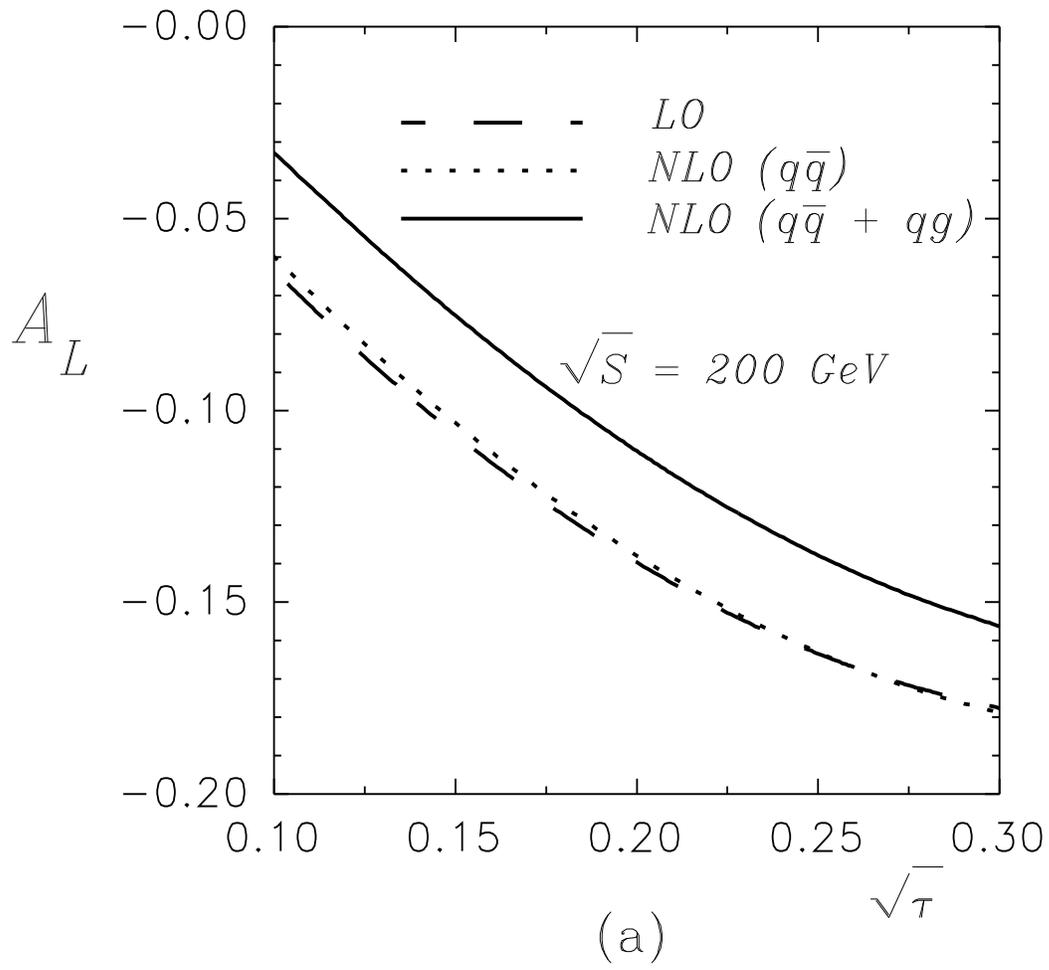
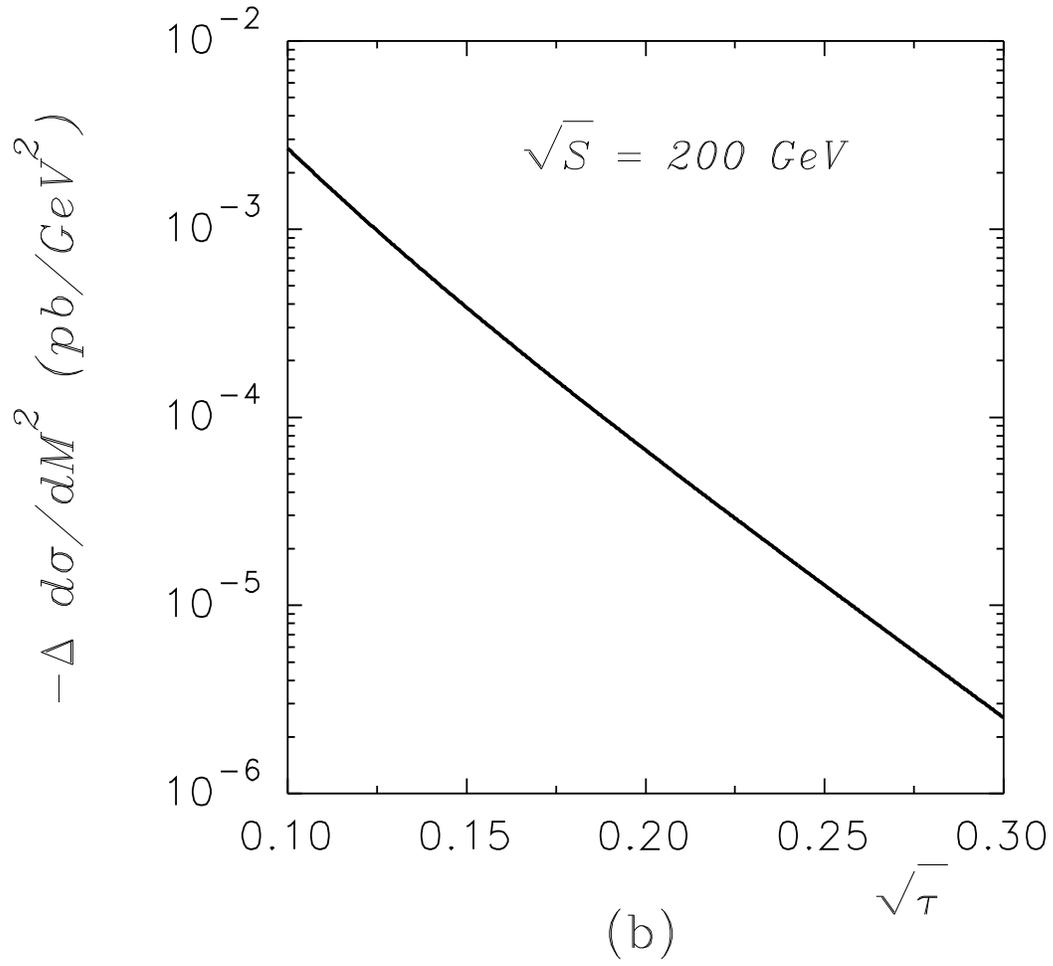
\end{document}